\documentclass[aps,prl,twocolumn,epsfig,amsmath,amssymb]{revtex4}
\usepackage[english]{babel} \usepackage{latexsym}
\usepackage{graphicx} \usepackage{subfigure} \usepackage{epsfig}
\usepackage{psfrag}

\newcommand{\re}{\mbox{Re}\,} 

\begin{document}

\title{Mott-insulator phase of coupled 1D atomic gases in a 2D optical
lattice} 
\author{D.M. Gangardt$^{1}$, P. Pedri$^{2,3,4}$,
L. Santos$^{2,3}$, and G.V. Shlyapnikov$^{1,5}$} 
\affiliation{$^1$\mbox{Laboratoire Physique Th\'eorique et Mod\`eles Statistiques, Universit\'e Paris Sud, B\^at. 100 91405 Orsay Cedex, France}\\
$^2$\mbox{Institut f\"ur Theoretische Physik III, Universit\"at
Stuttgart, Pfaffenwaldring 57, D-70550, Stuttgart, Germany}\\
$^3$\mbox{Institut f\"ur Theoretische Physik, Universit\"at Hannover, D-30167 Hannover, Germany}\\ 
$^4$\mbox{Dipartimento di Fisica, Universit\`a di Trento and BEC-INFM, I-38050 Povo, Italy}\\
$^5$\mbox{Van der Waals - Zeeman Institute, University of Amsterdam, Valckenierstraat 65/67, 1018 XE Amsterdam, The Netherlands}
}

\begin{abstract}  
%
%
We discuss the 2D Mott insulator (MI) state of a 2D array of coupled
finite size 1D Bose gases. It is shown that the momentum distribution
in the lattice plane is very sensitive to the interaction regime in
the 1D tubes. In particular, we find that the disappearance of the
interference pattern in time of flight experiments will not be a
signature of the MI phase, but a clear consequence of the strongly
interacting Tonks-Girardeau regime along the tubes.
\end{abstract}  
 \maketitle



Remarkable developments in atomic cooling and trapping
have triggered the interest in strongly
correlated atomic systems
\cite{biga,Gunn,Olshanii1,Recati,Jaksch98}. In particular, the physics
of cold atoms in periodic potentials induced by laser standing waves
(optical lattices) has attracted a major attention, mostly due to its
links to solid state physics \cite{BECLatt}, and due to the
observation of the superfluid (SF) to Mott-insulator (MI) transition
\cite{Fisher} in Munich experiments \cite{Greiner}. 


The reduction of spatial dimensionality in these systems is now a "hot topic",
in particular with regard to the creation of 1D gases, where 
the interaction between particles becomes
more important with decreasing the gas density. For low densities or
large repulsive interactions the system enters the strongly-interacting
Tonks-Girardeau (TG) regime, in which the bosons acquire fermionic
characteristics \cite{Girardeau60} and dynamical and correlation
properties drastically change \cite{Olshanii2,Chiara,Letter,Gangardt}.
This regime requires tight
transverse trapping, low atom numbers, and possibly the enhancement of
interactions via Feshbach resonances
\cite{Olshanii2,Olshanii1,Petrov1D}.  In this sense, 2D optical
lattices are 
favorable, since the on-site transverse
confinement can be made very strong and for sufficiently small
tunneling rate each lattice site behaves as an independent 1D system.
Recent experiments on strongly correlated 1D gases have been performed
along these lines \cite{Esslinger1,Esslinger2,Phillips,Bloch}.
 
These studies motivate the analysis of an interesting physics in a
(2D) array of coupled 1D Bose gases. In a 2D lattice the coupling is
provided by the inter-site tunneling, and each lattice site is a 1D
tube filled with bosonic atoms. This regime is easily achievable
experimentally by lowering the lattice potential, and it represents
the bosonic analog of 1D coupled nanostructures
\cite{Luttinger}.  As was first shown by Efetov and Larkin \cite{Efetov},  
for infinitely long 1D tubes at zero temperature 
any infinitesimally small tunneling drives the system into the
superfluid phase. The gas then enters an interesting
cross-dimensional regime, in which it presents 1D properties in a 3D
environment \cite{QuasiTonks,Cazalilla}. 
For 1D tubes of finite length $L$, at sufficiently small tunneling
rate the system can undergo a cross-over from such anisotropic 3D superfluid 
state to the 2D Mott insulator state \cite{Cazalilla}. Strictly speaking,
this 2D MI phase requires a commensurable filling of the tubes, i.e. an
integer average number of particles $N$ per tube. Then the 
system of finite-length tubes at zero temperature is analogous to that of
infinite tubes at a finite temperature $T$, and the critical tunneling $t_c$ for
the $T=0$ cross-over to the MI phase can be obtained from the finite-temperature
results of Ref. \cite{Efetov} by making a substitution $1/T\to L$ \cite{Rk1}.
 

This Letter is dedicated to the analysis of correlation properties of
this 2D Mott insulator.  We show that the momentum distribution is
crucially modified by a combined effect of correlations along the 1D
tubes and inter-tube hopping. For the case of a weakly interacting gas
in the tubes, the phase coherence is maintained well inside the MI
phase.  This is similar to the situation in 2D and 3D lattices, studied by
means of Quantum Monte Carlo calculations \cite{Proko} and
investigated experimentally through the observation of an interference
pattern after switching off the confining potential \cite{Greiner}.
However, an increase of the interaction between particles in 1D tubes
reduces the inter-tube phase coherence and flattens the momentum
distribution in the transverse direction(s).  In particular, the
interference pattern observed in Ref.~\cite{Greiner} should be largely
blurred if the 1D tubes are in the TG regime. This effect can be
revealed in current experiments. Remarkably, the disappearance of the
interference pattern will not be a signature of the MI phase, but a
clear consequence of the strongly interacting regime for 1D tubes.



In the following we consider a Bose gas at zero temperature in a 2D
optical lattice, such that every lattice site can be considered as an
axially homogeneous 1D tube of finite size $L$, with $N=nL$ being the
number of particles per tube, and $n$ the 1D density.  The tunneling
between neighboring tubes is characterized by the hopping $t$, which
depends on a particular lattice potential and atomic species
employed. We label tubes by the index $j$ and denote by $x$ the
coordinate along the tubes.

The action describing the coupled tubes has the form:
\begin{equation}
  \label{eq:S}
S=\sum_j S_j -t \sum_{<ij>}
  \int_{-\infty}^{\infty}\!\!\!d\tau\!\!\int_{-L/2}^{L/2}\!\!\! dx\;
  \left(\bar\psi_i\psi_j+\bar{\psi}_j\psi_i\right),
\end{equation}
where  $\psi_j (x,\tau),
\bar{\psi}_j (x,\tau)$ are complex bosonic fields associated with the
$j$-th tube, the symbol $<ij>$ denotes nearest neighbors, and $S_j$ is
the action describing the physics along the $j$-th tube.

In the absence of tunneling, there are no correlations between
different tubes, and the one-body Green function is diagonal:
\begin{eqnarray}
  \nonumber G_{ij}(x_1\!-\!x_2,\tau_1\!-\!\tau_2)
  &=&\langle\psi_i(x_1,\tau_1)\bar{\psi}_j(x_2,\tau_2) \rangle\\ &=&
  \delta_{ij} G_0(x_1\!-\!x_2,\tau_1\!-\!\tau_2).
  \label{eq:G}  
\end{eqnarray}
The presence of tunneling between neighboring tubes, provided by the
second term on the rhs of Eq. (\ref{eq:S}), modifies the momentum
distribution. Above a critical tunneling amplitude $t_c$
the system undergoes a cross-over from the MI to an anisotropic 3D SF
phase \cite{Cazalilla}.  This cross-over and the MI phase can be 
analyzed within the random phase approximation
(RPA) \cite{Tsvelik}, successfully used in the studies of coupled spin chains 
\cite{heisenberg}. Decoupling the tunneling term in the action (\ref{eq:S}) by using the
Hubbard-Stratonovich transformation and keeping only the leading
quadratic terms, yields the RPA Green function in the momentum-frequency representation:
\begin{equation}
  \label{eq:Gres}
    G (\vec q, k,\omega) = \frac{G_0(k,\omega)} {1-T(\vec q)
   G_0(k,\omega)},
\end{equation}
with $\vec{q}=(q_y,q_z)$ being the quasimomentum in the lattice plane,
$T(\vec q) = 2t(\cos q_y a+\cos q_z a)$, $a$ the lattice constant, and
$G_0 (k,\omega)$ the Fourier transform of the Green function
(\ref{eq:G}) (hereinafter we put $\hbar =1$): 
\begin{equation}
  \label{eq:G_FT}
  G_0 (k,\omega) = \int_{-\infty}^\infty d\tau\int_{-L/2}^{L/2}
  dx\,e^{-ikx+i\omega\tau} G_0 (x,\tau) .
\end{equation}

The long-wavelength behavior of the Green function $G_0(x,\tau)$ can be
found using Luttinger liquid theory 
\cite{Haldane1981}. At zero temperature, employing a conformal 
transformation in order to take into account the finite size $L$ of
the tubes \cite{Tsvelik}, we obtain:
\begin{equation}
\label{g0tfin}
  G_0(x,\tau) =n \left(\frac{\pi^2/N^2}{\sinh\left(\pi\zeta/L\right)
  \sinh\left(\pi\bar{\zeta}/L\right)} \right)^d,
\end{equation}
where $\zeta=v_s\tau+ix$, and $v_s$ is the sound velocity. The interactions 
enter Eq.(\ref{g0tfin}) through the factor $d=1/4K$ related to  the  interaction-dependent 
Luttinger parameter $K$. The Fourier transform of Eq.~(\ref{g0tfin}) yields
\begin{equation}
 \label{g0kfin}
  G_0(k,\omega) =\frac{1}{
  nv_s}\left(\frac{N}{2\pi}\right)^{2-2d} I\left(\frac{k
  L}{2\pi},\frac{\omega L}{2\pi v_s}\right),
\end{equation}
 where the quantity $I(p,\Omega)$ is expressed through the
 hypergeometric function $_3F_2$:


\begin{eqnarray}
\label{eq:Idef}
&&\!\!\!\!\!\!I(p,\Omega) =
\frac{4\pi}{p!}\frac{\Gamma(d+p)}{\Gamma(d)}\times
\nonumber\\ && \!\!\!\!\!\! \re \!\left[ \frac{
{}_3F_2\left(d,d\!+\!p,\frac{d+p-i\Omega}{2};
1\!+\!p,1\!+\!\frac{d+p-i\Omega}{2};1\right)} {d+p-i\Omega}
\right],
\end{eqnarray}
with $p=kL/2\pi$ and $\Omega=\omega L/2\pi v_s$ being the
dimensionless momentum and frequency. Integrating Eq.~(\ref{g0kfin})
over $\omega$ one obtains the momentum distribution
$N_0(kL/2\pi)=\int d\omega G_0 (k,\omega)/2\pi$ in the absence of
tunneling:
\begin{equation}
\label{eq:No_p}
  \frac{N_0(p)}{N} = \left(\frac{N}{2\pi}\right)^{-2d}\!
  \frac{\Gamma(d+p)}{p!\Gamma(d)}
  {}_2F_1(d,d+p;1+p;1),
\end{equation}
which behaves as $p^{2d-1}$ for $p\gtrsim 1$. The Luttinger
liquid description employed here is valid for low momenta $k\ll
\pi n$. Accordingly, the dimensionless axial momentum $p=kL/2\pi$, which
is an integer number, should satisfy the inequality $p\ll N$. The
momentum distribution (\ref{eq:No_p}) represents the fraction of
particles in the state with momentum $p$ and is normalized as $\sum_p
N_0 (p)=N$.


The critical tunneling $t_c$ for the MI to SF cross-over is obtained
as the value of $t$ for which the denominator of Eq.~(\ref{eq:Gres})
vanishes for zero momenta $k$ and $\vec{q}$ and zero frequency
$\omega$.  We thus have
\begin{equation} \label{tc}
\frac{t_c}{\mu} = \frac{n v_s}{4\mu}
\left(\frac{N}{2\pi}\right)^{2d-2} \frac{1}{I(0,0)}.
\end{equation}  
Note that with $t_c$ from Eq.~(\ref{tc}), the Green function $G$ in
Eq.~(\ref{eq:Gres}) becomes a universal function of the dimensionless
quantities $t/t_c$, $p$, $qa$ and $\Omega$.  For the TG regime of 1D
bosons in the tubes, the Luttinger parameter is $K=1$ and
$d=1/4$.  Then, as the chemical potential is
$\mu=mv_s^2=\pi^2n^2/2m$, from Eq.(\ref{tc}) we obtain
$t_c/\mu\simeq 0.05 N^{-3/2}$. For the weakly interacting regime, the
Luttinger parameter in the 1D tubes is $K=\pi(n/mg)^{1/2}\gg 1$
and Eq.(\ref{eq:Idef}) gives $I(0,0)=16\pi K\gg 1$.  In this regime
the chemical potential is $\mu=mv_s^2=ng$, and Eq.(\ref{tc}) then
yields $t_c/\mu\simeq (1/16) N^{-2}$ (as expected from the mean-field
calculations for a 2D lattice of zero-dimensional sites \cite{Stoof}).
These results are in qualitative agreement with the recent
calculations of Ho et al. \cite{Cazalilla}. One clearly sees that
strong correlations along the tubes drastically shift the boundaries
of the MI phase.

\begin{figure}[ht] 
\begin{center}\
\psfrag{y}{\hspace*{-1.0cm} $N(q,0)/N(0,0)$} 
\psfrag{x}{\hspace*{-0.4cm} $qa/\pi$}
\includegraphics[width=6.3cm]{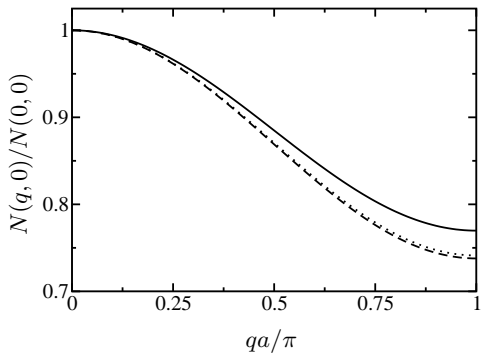}
\end{center} 
\caption{Transverse momentum distribution for $k=0$ at 
$t/t_c=0.3$ for $K=1$ (solid), $K=4$ (dotted), and $K=25$ (dashed).
In the figure we have chosen $q_y=q_z=q$.}
\label{fig:1}  
\end{figure}
\begin{figure}[ht] 
\begin{center}
\psfrag{y}{\hspace*{-1.0cm} $N_{\perp}(q)/N_{\perp}(0)$}
\psfrag{x}{}
\includegraphics[width=6.3cm]{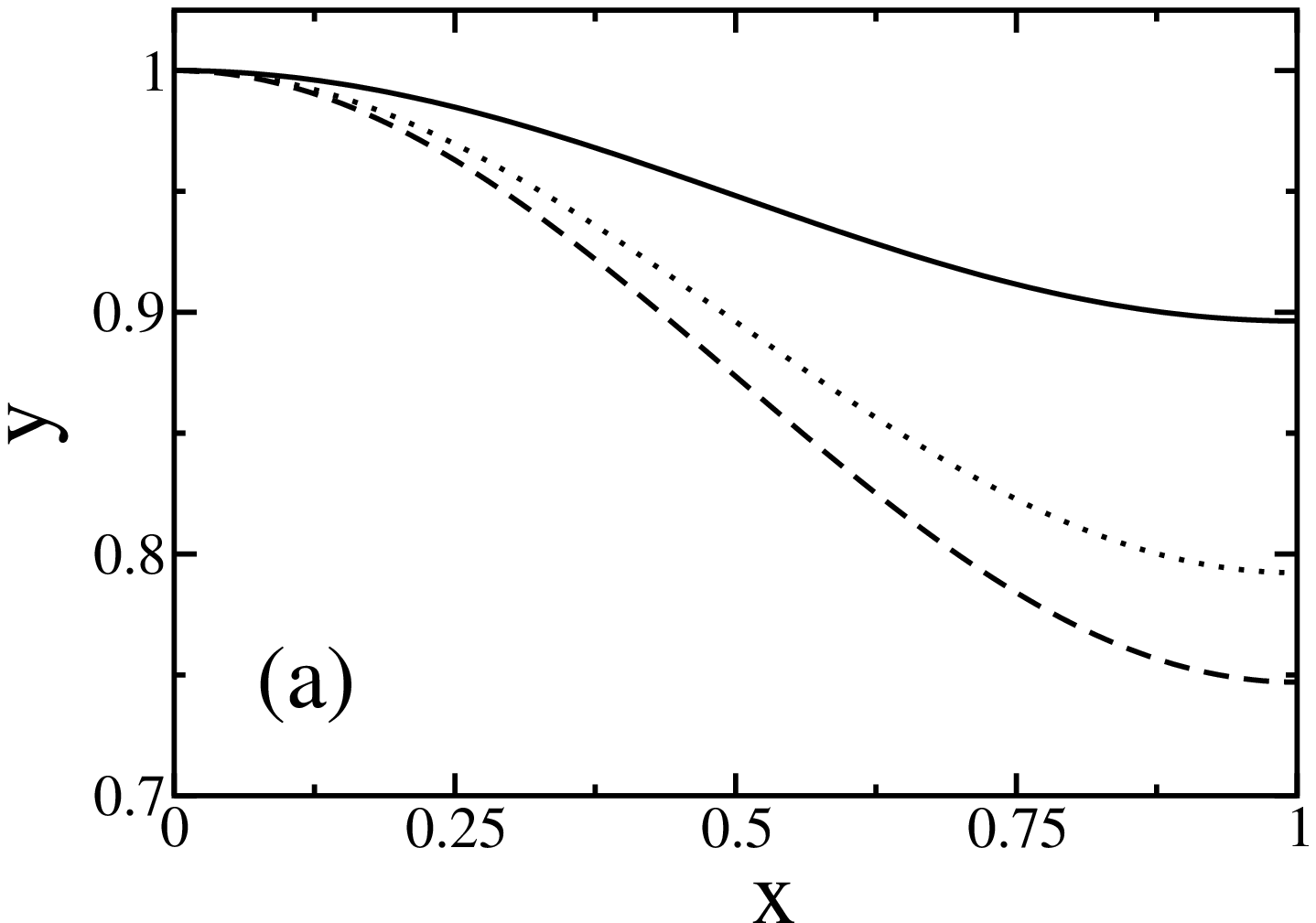}
\psfrag{x}{\hspace*{-0.4cm} $qa/\pi$}
\includegraphics[width=6.3cm]{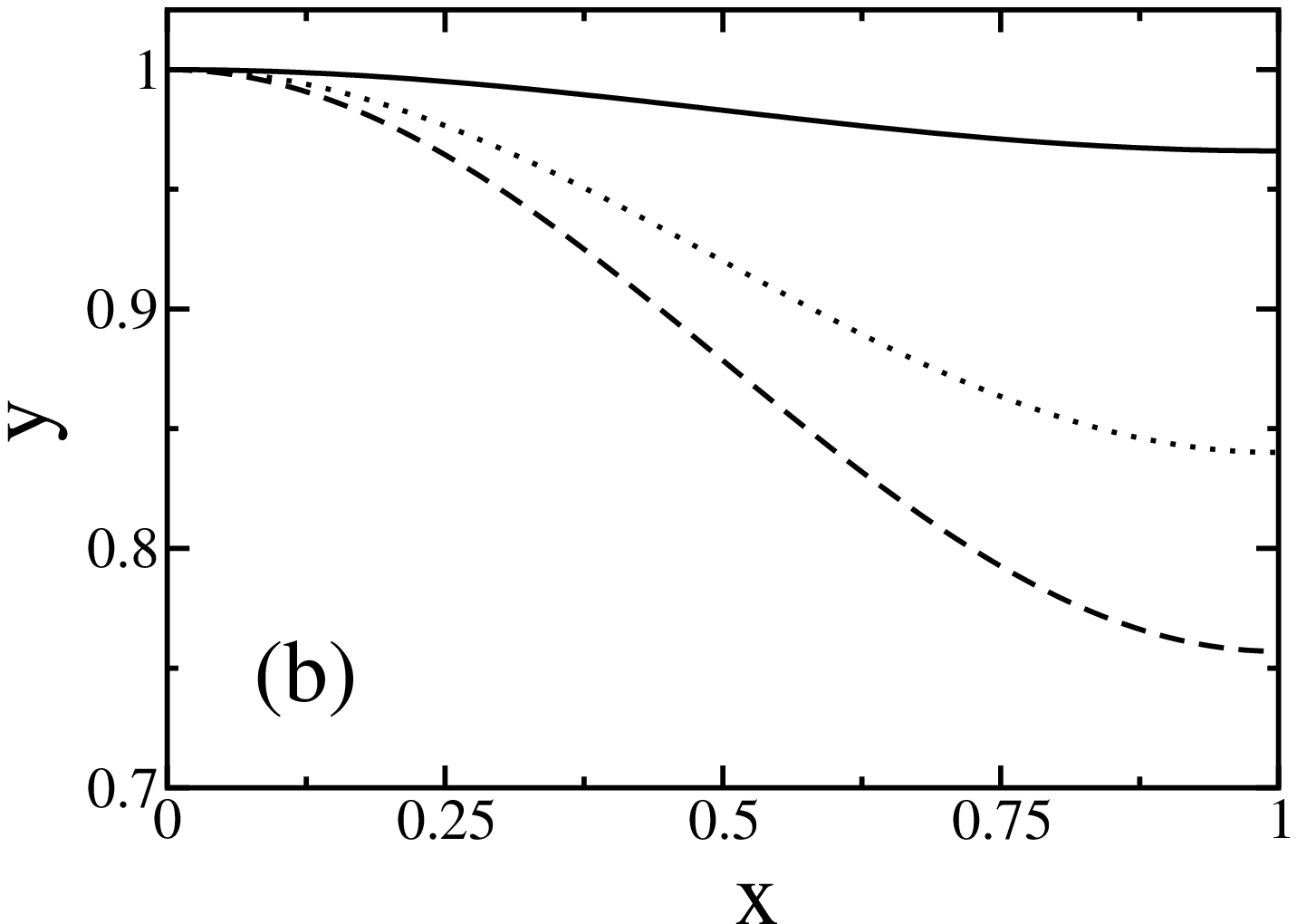}
\end{center} 
\caption{Transverse momentum distribution for $t/t_c=0.3$, 
at $K=1$ (solid), $K=4$ (dotted), and $K=25$ (dashed), for $N=50$ (a)
and $N=500$ (b).  In the figures we have chosen $q_y=q_z=q$.}
\label{fig:2}  
\end{figure}

The momentum distribution for the coupled 1D tubes in the MI phase,
$N(\vec{q},k)$, is obtained by integrating the Green function
(\ref{eq:Gres}) over the frequency.  
Our calculations show that only the lowest axial
mode for which the momentum $k=0$, is significantly affected by the
tunneling. The physical reason is that the $k=0$ 
mode is approaching the instability 
on approach to the critical tunneling $t_c$, whereas $k\neq 0$ modes are still
far from instability. This is reflected in the resonance character of the
Green function (\ref{eq:Gres}) for $k=0$ and $t\rightarrow t_c$. 
Our results in Fig.~\ref{fig:1} show that the transverse distribution
$N(\vec{q},0)$ corresponding to the $k=0$ axial mode is non-flat for
any interaction regime along the tubes.
In contrast, for $k\neq 0$ the quantity $T(\vec{q})G_0(k,\omega)$ is always
small. Therefore, expanding the rhs of Eq.~(\ref{eq:Gres}) in powers of $T(\vec{q})$ 
up to linear order and integrating over $\omega$ we obtain an almost flat
transverse momentum distribution for $k\neq 0$ modes: $N(\vec{q},p)/
N(0,p)=1-(Ad^2/4p^{2-d})(t/t_c)(2-\sum_{i=y,z}\cos q_ia)$, where the coefficient
$A$ is of order unity and the second ($q$-dependent) term is always very small.  

We now turn to the discussion of the transverse quasimomentum
distribution $N_\perp (\vec{q})=\sum_k N(\vec{q},k)$. The
summation over the axial modes changes the picture drastically
compared to the distribution for a given $k$. 
As only the $k=0$ component is significantly affected by the
tunneling, one can rewrite Eq. (\ref{eq:Gres}) in the form:
\begin{equation}\label{eq:G_approx}
G(\vec{q},k,\omega)\simeq G_0(k,\omega)+ \frac{T(\vec
q)G^2_0(0,\omega)} {1-T(\vec{q})G_0(0,\omega)}\delta_{k,0} .
\end{equation}
In the second term on the rhs of Eq.(\ref{eq:G_approx}) we may use the
Green function $G_0(0,\omega)$ following from Eqs.~(\ref{g0kfin}) and
(\ref{eq:Idef}).  For $k=0$ ($p=0$), one can put the hypergeometric
function $_3F_2=1$ in Eq.(\ref{eq:Idef}), which gives $I
(0,\Omega)\simeq 4\pi d/(d^2+\Omega^2)$.  Omitted terms give
a very small relative correction of the order of $d^3<1/4^{3}$. 
Hence, using Eq.(\ref{tc}), for the Green function
at $k=0$ in the absence of tunneling we have $G_0(0,\omega)
=d^2/4t_c(d^2+\Omega^2)$.  Then, integrating
Eq.(\ref{eq:G_approx}) over $\Omega$, summing over the axial modes
$k$, and imposing the normalization condition $N=\sum_k\int d\omega
G_0 (k,\omega)/2\pi$ for the first term on the rhs, we obtain the
transverse momentum distribution
\begin{equation}
\label{bure}
\!\!\frac{N_\perp(\vec{q})}{N}\!=\!1\!+\!
\left(\frac{2\pi}{N}\!\right)^{2d}\!\!
\left[\!\left(\!1\!-\!\frac{t}{2t_c}\sum_{i=y,z}\!\!\cos{q_ia}\right)^{\!-1/2}
\!\!\!\!\!\!\!\!-\!1\right]\!,\!\!
\end{equation}
normalized by the condition $(a/2\pi)^2\int d^2qN_{\perp}(\vec{q})=N$.

In Fig.~\ref{fig:2} we depict the results of Eq.~(\ref{bure}) for
different values of $N$ and the Luttinger parameter $K$.  
Due to the prefactor in the second term on the rhs of Eq.~(\ref{bure})
the transverse momentum distribution strongly
depends on the interaction regime along the tubes.
 
For the weakly interacting regime ($d\ll 1$), the distribution
$N_\perp(\vec{q})$ is not flat even deeply inside the MI phase.
Similar results have been obtained by means of Quantum Monte Carlo
calculations \cite{Proko} for the case of 
lattices of zero-dimensional sites.  In our case, only for rather low
tunneling ($t/t_c \lesssim 0.1$) the quasimomentum distribution
becomes flat, and switching off the lattice potential should lead to a
blurred picture as that observed by Greiner et al. \cite{Greiner}.
Non-flat distributions in the MI phase as those of Fig.~\ref{fig:2}
will manifest themselves through the appearance of interference peaks
in the same type of experiment.

On approach to the TG regime, the quasimomentum distribution becomes
progressively flatter.  The main reason for this behavior is that when
the system becomes more interacting, the $k=0$ component is more
depleted, contributing less to the total
quasimomentum distribution.  Therefore, if the 1D tubes approach the
strongly interacting TG regime, in experiments as those of
Ref.~\cite{Greiner} the interference pattern will be essentially
smeared out.  One may expect a partial destruction of the interference
pattern even for moderate values of the Luttinger parameter (see,
e.g., the case $K=4$ in Fig.~\ref{fig:2}).

The random phase approximation used in our calculations, was shown to 
be a good approximation for a wide range of parameters of coupled 
one-dimensional Heisenberg spin chains \cite{heisenberg}.  Here we give 
yet another estimate for the applicability of RPA, relying on the
Ginzburg criterion adapted to a quantum phase transition at zero temperature. 
We compare fluctuations of  the order parameter in a volume 
determined by the correlation radius extracted from the Green function
(\ref{eq:Gres}), with the scale on which the non-linear 
effects become important. The latter is obtained from the four-point
correlation function of each tube. We have found that RPA is adequate for
$(t_c-t)/t_c\gg B(d)$, where $B(d)$ has been  obtained numerically
from the  four-point correlation function. In the case of the 
Tonks-Girardeau regime along  the tubes, we have $B\approx 0.1$, and it 
decreases significantly with decreasing $d$ and entering the  
Gross-Pitaevskii regime.


In conclusion, we have considered a bosonic gas in a 2D optical
lattice of finite 1D tubes at zero temperature, focusing our attention
on the momentum distribution in the MI phase.  We have shown that the
strong correlations along the 1D tubes significantly modify the
quasi-momentum distribution in the lattice plane.  We have found that
in the MI regime only the lowest momentum along the tubes is affected
by the inter-site hopping, and hence only this component contributes
to the formation of interference fringes. Consequently, the larger the
interactions are (larger depletion) the less pronounced is the
visibility of the interference fringes.  In particular, for the TG
regime in the tubes, the quasimomentum distribution becomes
progressively flatter, leading to an observable blurring of the
interference pattern after expansion.  This effect can be observed in
current time of flight experiments, and can be used to reveal a clear
signature of the strong correlations along the sites.

We acknowledge discussions with M.A. Cazalilla, A.F. Ho, and with the group of 
A. Muramatsu. This work was supported by the Centre National de la Recherche
Scientifique (CNRS), the Nederlandse Stichtung voor Fundamenteel
Onderzoek der Materie (FOM), Deutsche Forschungsgemeinschaft SFB 407
and SPP1116, the RTN Cold Quantum Gases, IST Program EQUIP, ESF PESC
BEC2000+, EPSRC, the Ministero dell'Istruzione, dell'Universit\`a e
della Ricerca (MIUR), the Alexander von Humboldt Foundation, and in part by 
the National Science Foundation under Grant No. PHY99-07949. 
LPTMS is a mixed research unit No. 8626 of CNRS and Universit\'e Paris Sud.


\begin{thebibliography}{99}
\bibitem{Rk1} We assume that the commensurability condition is
satisfied. Note, however, that for a large non-integer $N$ one has only a tiny 
superfluid fraction ($\sim 1/2N$ or less), and below $t_c$ the 
system behaves in almost all aspects as the MI phase.
\end{thebibliography}
\end{document}